\documentclass{ws-brl}
\begin{document}

\markboth{Hector Socas-Navarro}
{On the connection between planets, dark matter and cancer}

\catchline{}{}{}{}{}

\title{ON THE CONNECTION BETWEEN PLANETS, DARK MATTER AND CANCER}

\author{Hector Socas-Navarro}

\address{Instituto de Astrof\'\i sica de Canarias\\
  Avda V\'\i a L\'actea S/N, La Laguna 38205, Tenerife 38205, Spain}
\address{Departamento de Astrof\'\i sica, Universidad de La Laguna\\
  La Laguna, Tenerife 38205, Spain\\
hsocas@iac.es}

%

\maketitle

\begin{history}
\received{Day Month Year}
\revised{Day Month Year}
\end{history}

\begin{abstract}
  In a recent paper, Zioutas and Valachovic (2018) claim that dark
  matter is responsible for a significant fraction of the melanoma
  skin cancer. This conclusion is drawn from their observation of a
  significant correlation between skin melanoma incidence in the US
  and the inner planets positions (especially those of Mercury and
  Earth).  Here I present a number of objections to their
  interpretation. Some (but not all) of the counterarguments are based
  on the analysis of a larger dataset from the same source,
  considering more cancer types and separating by patient attributes,
  such as race. One of the counterarguments is that, if the melanoma
  fluctuations with periods similar to planetary orbits were produced
  by dark matter density enhancements on Earth, then we would have to
  conclude that the black population is somehow immune to dark matter,
  a conclusion that seems incompatible with the current WIMP
  paradigm. Interestingly, some periodicities are present in the data,
  including the ones reported by Zioutas and Valachovic, but I argue
  that they must have a societal rather than astronomical origin.

\end{abstract}

\section{Introduction}

In the most widely accepted cosmological model (the $\Lambda$CDM
cosmology, named after Einstein's cosmological constant $\Lambda$ and
the Cold Dark Matter acronym), dark matter is an invisible substance
that accounts for nearly 80\% of all matter in the
Universe\cite{Planck18}. It is not affected by electro-magnetic forces
and, as such, cannot be seen or touched, which has allowed this
enigmatic form of matter to elude direct detection thus far. A vast
array of astrophysical observations during the last decades has
resulted in strong constraints on its properties, leading to a
majority consensus among experts in favor of the WIMP
(Weakly-Interacting Massive Particles)
scenario\cite{RST18}. Basically, dark matter would be a fluid of
subatomic particles interacting only by gravity and by the weak
nuclear force. In this respect, the dark matter particles would be
similar to neutrinos, except that neutrinos move at nearly the speed
of light whereas the dark matter particles inferred by galactic and
cosmological observations must be moving at much lower,
non-relativistic speeds (this is the meaning of ``cold'' in the CDM
acronym). The dark matter fluid should be in the form of nearly
spherical halos, in which galaxies are embedded. In our Milky Way,
dark matter particles in the solar neighborhood should be moving at
speeds of the order of 230~km/s. Being electrically neutral, dark
matter particles do not come together to form molecules and do not
collapse to form disks, stars or planets, like ordinary matter
does. In the ideal case of an isolated galaxy, a dark matter halo
would have a smooth, spherically-symmetric density distribution,
without structure at scales much smaller than the galaxy
size\cite{NFW96}. However, tidal interactions or gravitational
perturbations from nearby objects may induce substructure in the
galactic halos, such as the so-called streams. While observational
evidence of such substructures has not yet been found, it is being
actively sought by astronomers.

Because dark matter has a rather uniform distribution at interstellar
scales, it has a relatively low density compared to ordinary matter in
the vicinity of stars. This is because ordinary matter is extremely
concentrated in and around stars, separated by colossal distances of empty
space between them. The amount of dark matter mass contained in the
solar system is estimated to be comparable to that of a large
asteroid. This is the reason why dark matter is only important at
galactic scales or larger.

Every second, a number of dark matter particles, moving at speeds of
hundreds of km/s, pass through our bodies. The actual number depends
on the WIMP mass (or mass distribution), which is unknown. Since these
particles could, in principle, collide and interact with the atomic
nuclei in our cells, they might be able to alter the DNA composition
and induce mutations in our genes, thus increasing the risk for
developing cancerous tumors. Given our present lack of knowledge on
the WIMP properties, it is not possible to estimate the rates of
collisions or mutations produced by these
particles\cite{FS12}\cdash\cite{Z90}. However, as the solar system
moves through the galaxy, it is expected to encounter different dark
matter densities on scales from tens to hundreds of millions of
years. Some authors have sought indirect evidence of dark matter
interaction fluctuations in the geological or fossil
records\cite{Randall15}\cdash\cite{Rampino15}. Connections between
cancer rate variations and dark matter density have been proposed in
the literature.

Zioutas and Valachovic\cite{ZV18} (hereafter ZV18) claimed to have found a
link between periodic variations of skin melanoma and dark matter
density changes driven by planetary motions in the solar
system. Their paper is not the first to put forward a causal relation
between dark matter and cancer but it is the first to correlate it
with planetary motions and on timescales of months, as opposed to at
least tens of millions of years in previous works.

In the spirit of healthy scientific debate, this paper puts forward
some objections to the conclusions of ZV18 that should probably be
kept in mind in future studies on the issue which, undoubtedly, is an
important one. As they point out, cancer and the ``Dark Universe'' are
among the biggest mysteries in medicine and physics, respectively. The
following sections present first an introduction to the cancer data
analyzed here and then the counterarguments to the planetary link
hypothesis in ascending order of strength.

\section{Cancer database}

ZV18 employed for their analysis the incidence of skin melanoma
compiled in the database SEER9 from the Surveillance Epidemiology End
Results of the US National Cancer Institute. This dataset comprises
records of individual patients diagnosed between 1973 and 2015 in a
sample population from Atlanta, Connecticut, Detroit, Hawaii, Iowa,
New Mexico, San Francisco-Oakland, Seattle-Puget Sound and Utah. The
database is organized by eight major cancer categories (breast, colon
and rectum, digestive, female genital, male genital, lymphoma and
leukemia, respiratory and urinary) and a miscellaneous category that
includes several other types, including skin melanoma. The data
presented in this paper are from the same database, although some
results discussed below have been obtained from the analysis of other
types of cancer.

\section{Counterarguments to the planetary link}
\subsection{Dark matter structure}

The astrophysical scenario considered by ZV18 is not described in
their paper. They assume the presence of a dark matter stream running
through the solar system. The correlation between planet positions and
skin melanoma is interpreted as a result of the planets periodically
increasing the density of dark matter passing through Earth. However,
this overall picture is very difficult, if not impossible, to
translate into a detailed layout. Consider Mercury, for instance. Let
us assume that there is a dark matter stream and Mercury moves into
and out of the stream regularly during its orbit. Furthermore, in
order to produce the observed periodicity, this geometry (and the
position at which Mercury dives into the stream) must remain the same
for the entire period of 38 years analyzed by ZV18. The Sun revolves
around the galaxy at approximately 200 km/s, which means that it has
traveled 2.4$\times$10$^{12}$~km in 38 years. In contrast, Mercury's
orbit is only 58 million km. This mismatch of several orders of
magnitude between both distances makes it nearly impossible to
construct a suitable geometrical scenario unless the dark matter
stream is perfectly aligned with the solar system motion through the
galaxy, to a precision better than 10$^{-4}$ radians (a percent of a
degree).

\subsection{Relative planet positions}

Even if Mercury and Venus were periodically diving into a stream,
focusing or splashing dark matter around, the position of Earth around
its orbit would be different each time they do this. In other words,
for Mercury to produce repeated enhancements of dark matter density at
Earth, one would need to replicate the relative positions of Mercury,
Earth and the stream. We would not have a periodicity with Mercury's
orbital period because every time Mercury encounters the dark matter
stream, Earth would be at different location.

\subsection{Diagnosis delay}

The dates recorded in the SEER database reflect the times when the
cancer was diagnosed, not when the patients became sick, much less
when they were exposed to a possible causing carcinogenic
agent. Cancer diagnosis delay is an extremely important parameter in
medicine, given its influence on the survival rate, and numerous
studies exist in the scientific literature. Such studies usually
consider the time from the patient noticing the first symptoms to
consultation with a primary care physician and then the further delay
to be referred and diagnosed by a specialist. Existing studies for the
particular case of skin melanoma conclude that the delay from symptoms
to diagnosis is typically of the order of 11 months\cite{KIH+91}
or, more recently, 7 months\cite{XDLB+16}. In the context of
timing a possible dark matter excess as a potential carcinogen, one
would need to consider an additional term, namely the delay from
exposure to the appearance of the first symptoms which, to the
author's knowledge, has not been quantified. Diagnosis delay is then
an average of 2 to 4 times longer than Mercury's orbit, which means
that, even if there were a causal connection between Mercury's
position and the incidence of melanoma, it would be completely washed
out in the data by the effect of diagnosis delay. A similar statement
could be made for Venus or Earth, with somewhat lower ratios of delay
to claimed periodicity, down to between 0.6 and 0.9 for the case of
Earth.

\subsection{Periodicities}

\begin{figure}[ph]
\centerline{\psfig{file=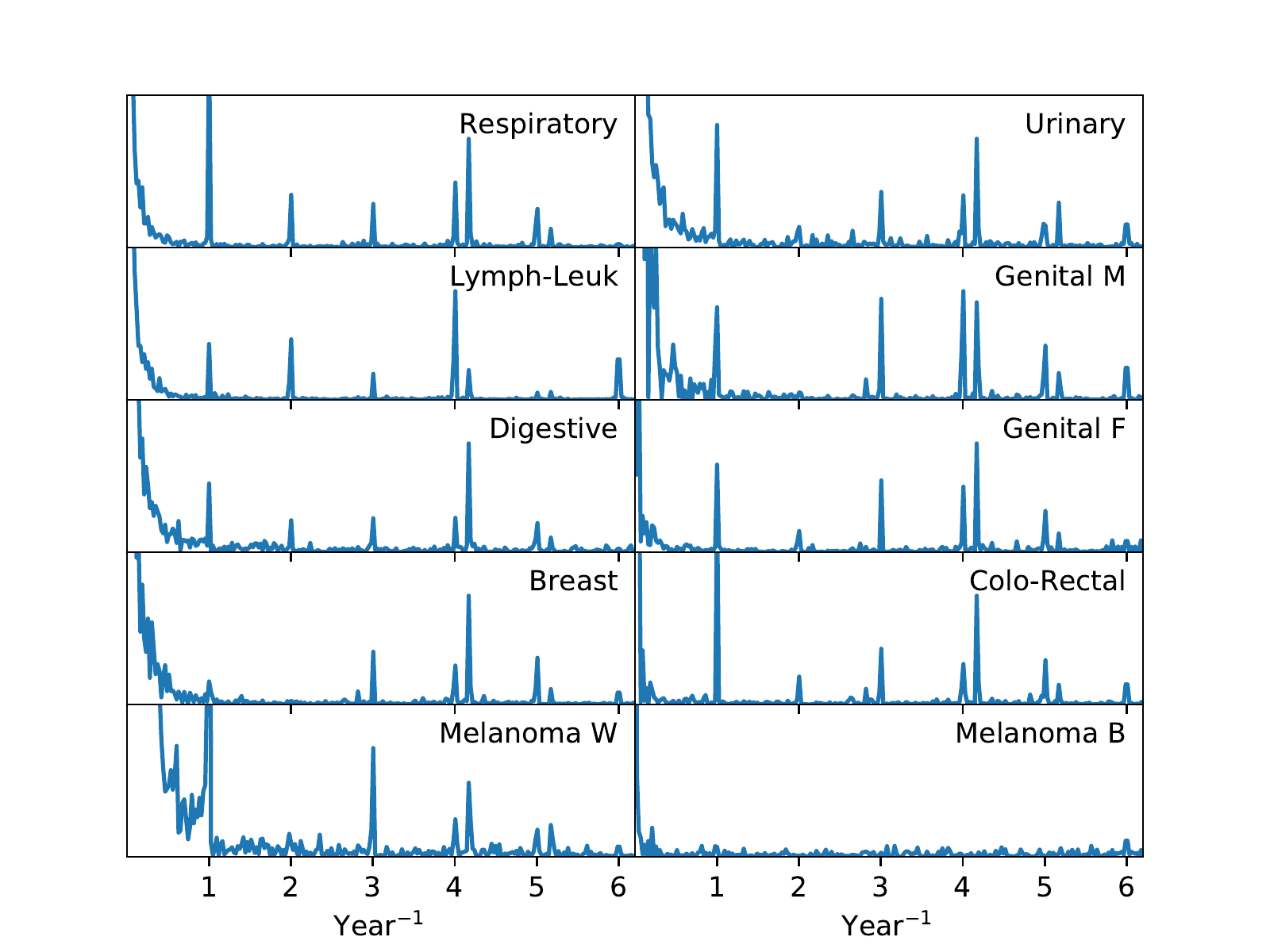,width=\textwidth}}
\vspace*{8pt}
\caption{Fourier power spectra of the 8 major cancer categories in the
  SEER9 database, along with the skin melanoma for white and black
  populations. Incidence is per month (normalized to 30.44 days) and
  100,000 population. The ``Lymph-Leuk'' label refers to ``lymphoma
  and leukemia''. ``Genital M'' and ``Genital F'' refers to male and
  female genital cancer, respectively.
 \label{ffts}}
\end{figure}

Although ZV18 restricted their analysis to the case of skin melanoma,
the SEER database includes many other types of
cancer. Figure~\ref{ffts} shows the Fourier power spectrum of the
eight major categories in the database alongside the skin melanoma,
which is displayed separately for white and black populations.

The power spectrum exhibits a number of conspicuous peaks above the
noise, indicative of periodicities in the rate of cancer
diagnosis. The most prominent peak occurs at one year. A simple
explanation for this periodicity is that it is caused by human habits
such as routine medical examinations, which are typically scheduled on
a yearly basis. The other peaks at integer positions are easily
explained as harmonics of the one-year peak. The Fourier decomposition
of a simple harmonic function (a sine or cosine signal) results in a
single peak at the signal frequency. However, if we have a function
that is periodic but not sinusoidal, we obtain a peak at the
fundamental frequency plus a number of other peaks at the multiples of
that frequency (harmonics). The relative amplitudes of the harmonics
depend on the specific shape of the function. Therefore, all the
peaks at integer abscissae values in Fig.~\ref{ffts} may be ascribed
to a yearly variation in cancer diagnosis rate, which is more frequent
in the summer of the northern hemisphere\footnote{It would be
  interesting to look for a similar pattern in southern hemisphere
  data.}.

Two peaks in the power spectra remain unaccounted for. Their
frequencies and widths are 4.17 $\pm$ 0.01 and 5.17 $\pm$ 0.01
year$^{-1}$, respectively. ZV18 associated the first one with
Mercury's orbit, which is of 88 days or 4.15 year$^{-1}$. They do not
give any explanation for the peak at 5.17. The analysis presented here
does not show any feature with the frequency of Venus orbit at 1.62
years$^{-1}$ (225 days).

Figure~\ref{ffts} shows that all of the main cancer categories in the
SEER database exhibit the same periodicities, the peaks are located at
the same frequency positions. However, the ZV18 paper presents a
connection of dark matter to skin melanoma only. The reasons for this
are not clear. They do not mention whether they analyzed the other
data series or what was the motivation to focus on skin melanoma in
particular.

\subsection{Race immunity}

An interesting observation from Figure~\ref{ffts} is the absence of
skin melanoma among the black population (bottom-right panel). Skin
pigmentation is a well-known protective factor against ultraviolet
radiation and skin cancer development. However, there is no known
reason why it should protect against dark matter. In order to explore
this possibility, we need to isolate the contribution of dark matter,
separating it from the rest of the white population melanoma
incidence. If, as claimed by ZV18, the 4.17 year$^{-1}$ peak is due to
dark matter density fluctuations induced by Mercury, we can filter out
that frequency and analyze the two contributions separately. The
filtered signal (first component) would be the cancer incidence due to
normal (non-dark matter related) causes and may be different for
different races. The remaining (second component) would account for
those cases that have been induced by dark matter. A simple filtering
would give us a fluctuating second component that oscillates between a
negative and a positive value, corresponding to the hypothetical
fluctuation in cases due to dark matter. We are interested in the
amplitude of this fluctuation.

\begin{figure}[ph]
\centerline{\psfig{file=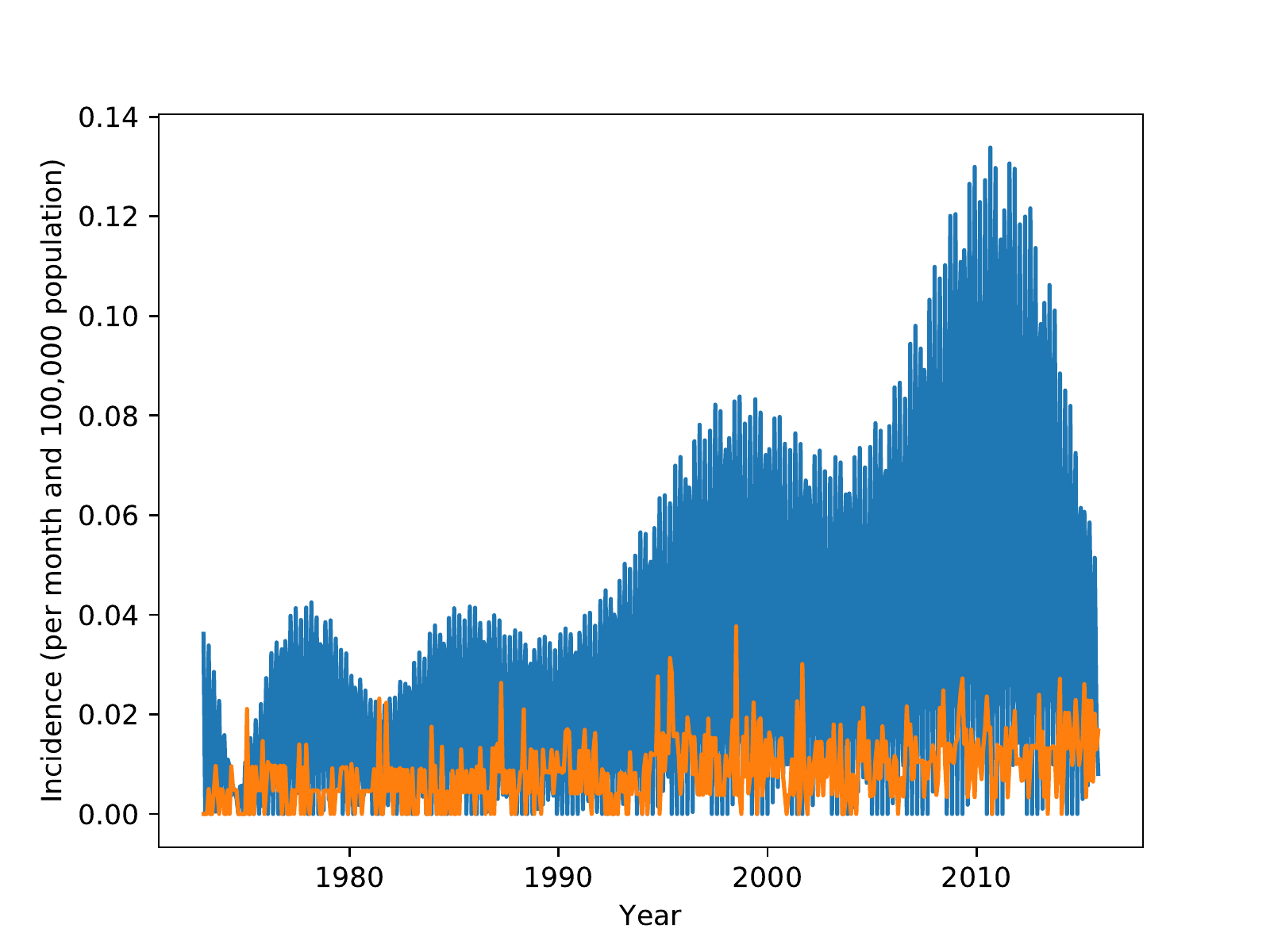,width=0.8\textwidth}}
\vspace*{8pt}
\caption{Blue: Time variation of melanoma skin cancer incidence
  associated to the 4.17 year$^{-1}$ peak in the white
  population. Orange: Total cancer incidence for the black
  population. Note that both curves are normalized by time and
  population.
 \label{bw}}
\end{figure}

Figure~\ref{bw} (blue) shows the second component envelope as a
function of time, i.e. the variation of the skin melanoma incidence in
the white population supposedly induced by dark matter. At the very
least, the total incidence of skin melanoma in the black population
should have a comparable amplitude to the blue curve. However,
Fig.~\ref{bw} shows that the total incidence for the black population
(orange) is always much smaller than the second component in the white
population. This means that, in the ZV18 hypothesis (that Mercury's
orbit is periodically enhancing the incidence of skin melanoma with an
88-day period), this enhancement does not affect the black population,
which would be difficult to explain in a model where dark matter does
not have electro-magnetic interaction (by definition) and can only
interact with atomic nuclei.

\section{Conclusions}	

The paper of ZV18 put forward a very bold claim which, if true, would
have a tremendous impact on physics and physiology, namely that there
is a causal connection between dark matter, planetary motions and
melanoma skin cancer. In particular, they highlight a link
with the motion of Mercury, which is by far the clearest signal
according to their analysis. This work presents a number of objections
to that claim. It does not rule out a possible relationship between
dark matter and cancer but, if such relationship exists, it probably
cannot be derived from these data.

It is possible that dark matter might form structures of galactic
scale and, in fact, we have identified streams of stars where wisps of
dark matter probably exist. Furthermore, it is possible that the solar
system moves through such structures, probing a varying dark matter
density. The most obvious example is the S1 stream in which the solar
system is actually embedded\cite{OHCMC+18}. However, the time scales
involved in such transits are much longer. For instance the Sun has
been moving through the S1 stream for roughly 5 million years and will
continue to do so for a similar amount of time. Another dark matter
structure that has been hypothesized is a galactic disk, assuming that
dark matter exhibits some (extremely weak) degree of interaction with
itself. The oscillatory motion of the solar system across the galactic
plane would make it plunge into and out of such disk but that would
occur on periods of about 60 million years. However, on planetary
motion scales, and particularly on scales of less than three months,
it is not expected to find such variations. There is no plausible
astrophysical scenario that would produce a splash of dark matter as
Mercury moves in and out of it consistently for over 40 years. This is
exacerbated if we consider that the relative positions of Mercury and
Earth are changing continuously. There is no way to reproduce a
geometry between Earth, Mercury and the putative dark matter source
with the periodicity of Mercury's orbit.

Even if we forget about all of the above considerations, there is a
very important problem with the ZV18 interpretation. The data that we
have do not reflect the time when the disease was contracted or even
when it manifests its first symptoms. We have the date of
diagnosis. Diagnosis delay is an important problem in medicine and it
would smear out any possible periodic signal that could have existed,
such as the putative planetary connection. The causes for any
periodicities in the data should be sought first in how we approach
the search and discovery of cancer in patients. The one-year period
and its harmonics found in most types of cancer are probably a direct
consequence of our medical examination habits. Interestingly, there
are two periodicities that, to the author's best knowledge, remain
unexplained. These peaks are located at 4.17 and 5.17 year$^{-1}$ (87
and 70 days).

Even if we forget about astrophysics and we also forget about
diagnosis delay, the data shows that the black population is
unaffected by the dark matter induced modulation that ZV18 claim to
see in the overall incidence of skin melanoma. Skin pigmentation
protects against the harmful effects of ultraviolet radiation and it
is less prone to cancer development but there is no known reason why a
darker skin should make people immune to dark matter.

From all of the above considerations, it appears that the link between
Mercury's orbital period and a periodicity in cancer (of many types)
diagnosis is casual and not causal. No evidence is observed in this
analysis for a similar coincidence with Venus.

\section*{Acknowledgments}

The author gratefully acknowledges financial support from the Spanish
Ministry of Economy and Competitivity through project
AYA2014-60476-P. This research has made use of NASA's Astrophysics
Data System Bibliographic Services.  The Python Matplotlib\cite{H07},
Numpy\cite{numpy11} and iPython \cite{ipython07} modules have
been employed to generate the figures and calculations in this paper.

\section{References}

  \bibliography{paper}



\end{document}